\documentclass[twocolumn,superscriptaddress]{revtex4}
\usepackage{amsfonts}
\usepackage{amsmath}
\usepackage{amsthm}
\usepackage{amscd}
\usepackage{amssymb}
\usepackage{subfigure}
\usepackage{amsxtra}
\usepackage{gensymb}
\usepackage{bm}           
\usepackage{bbm}
\usepackage{graphicx}
\usepackage{epstopdf}
\usepackage{color}

\def\bi#1\ei {\begin{itemize}#1\end{itemize}}
\def\bn#1\en {\begin{enumerate}#1\end{enumerate}}
\def\bea#1\eea {\begin{align}#1\end{align}}
\def\bean#1\eean {\begin{align*}#1\end{align*}}
\def\ben#1\een {\begin{equation*}#1\end{equation*}}
\def\be#1\ee {\begin{equation}#1\end{equation}}
\def\bes#1\ees {\begin{equation}\begin{split}#1\end{split}\end{equation}}
\def\bear#1\eear {\begin{eqnarray}#1\end{eqnarray}}
\def\bear#1\eear {\begin{eqnarray*}#1\end{eqnarray*}}

\newcommand{\beq}{\begin{equation}}
\newcommand{\eeq}{\end{equation}}

\newcommand{\ket}[1]{\ensuremath{\left|#1\right\rangle}}

\newtheorem{thm}{Theorem}

\begin{document}

\title{Experimental Quantum Fingerprinting}

\author{Feihu Xu}
\thanks{These authors contributed equally to this work.}
\address{Center for Quantum Information and Quantum Control, Department of  Electrical and Computer Engineering and Department of Physics, University of Toronto, Toronto, Ontario, M5S 3G4, Canada}
\address{Present address: Research Laboratory of Electronics, Massachusetts Institute of Technology, 77 Massachusetts Avenue, Cambridge, Massachusetts 02139, USA}
\author{Juan Miguel Arrazola}
\thanks{These authors contributed equally to this work.}
\address{Institute for Quantum Computing and Department of Physics and Astronomy, University of Waterloo, 200 University Avenue West, Waterloo, Ontario, N2L 3G1, Canada }
\author{Kejin Wei}
\address{Center for Quantum Information and Quantum Control, Department of  Electrical and Computer Engineering and Department of Physics, University of Toronto, Toronto, Ontario, M5S 3G4, Canada}
\address{School of Science and State Key Laboratory of Information Photonics and Optical Communications, Beijing University of Posts and Telecommunications, Beijing 100876}
\author{Wenyuan Wang}
\address{Center for Quantum Information and Quantum Control, Department of  Electrical and Computer Engineering and Department of Physics, University of Toronto, Toronto, Ontario, M5S 3G4, Canada}
\address{Department of Physics, University of Hong Kong, Pokfulam Road, Hong Kong, China}
\author{Pablo Palacios-Avila}
\address{Institute for Quantum Computing and Department of Physics and Astronomy, University of Waterloo, 200 University Avenue West, Waterloo, Ontario, N2L 3G1, Canada }
\address{Facultad de Ciencias, Universidad Nacional de Ingenieria, Lima, Peru.}
\author{Chen Feng}
\address{School of Engineering, University of British Columbia, Kelowna, British Columbia, V1V 1V7, Canada}
\author{Shihan Sajeed}
\address{Institute for Quantum Computing and Department of Electrical and Computer Engineering, University of Waterloo, 200 University Avenue West, Waterloo, Ontario, N2L 3G1, Canada }
\author{Norbert L\"{u}tkenhaus}
\email{lutkenhaus.office@uwaterloo.ca}
\address{Institute for Quantum Computing and Department of Physics and Astronomy, University of Waterloo, 200 University Avenue West, Waterloo, Ontario, N2L 3G1, Canada }
\author{Hoi-Kwong Lo}
\email{hklo@comm.utoronto.ca}
\address{Center for Quantum Information and Quantum Control, Department of  Electrical and Computer Engineering and Department of Physics, University of Toronto, Toronto, Ontario, M5S 3G4, Canada}

\date{\today}

\begin{abstract}
Quantum communication holds the promise of creating disruptive technologies that will play an essential role in future communication networks. For example, the study of quantum communication complexity has shown that quantum communication allows exponential reductions in the information that must be transmitted to solve distributed computational tasks. Recently, protocols that realize this advantage using optical implementations have been proposed. Here we report a proof of concept experimental demonstration of a quantum fingerprinting system that is capable of transmitting less information than the best known classical protocol. Our implementation is based on a modified version of a commercial quantum key distribution system using off-the-shelf optical components over telecom wavelengths, and is practical for messages as large as 100 Mbits, even in the presence of experimental imperfections. Our results provide a first step in the development of experimental quantum communication complexity.
\end{abstract}

\maketitle
What technological advantages can be achieved by directly harnessing the quantum-mechanical properties of physical systems? In the context of communications, it is known that quantum mechanics enables several remarkable improvements, such as cryptographic protocols that are classically impossible \cite{bennett84a,ekert91a,QDS}, enhanced metrology schemes \cite{giovannetti2011advances}, and reductions in the communication required between distributed computing devices \cite{Yao1979,BrassardQCC,RevModPhys.82.665,buhrman1999multiparty,buhrman1998quantum,QuantumFingerprinting,RazProblem,HM-Bar-Yossef,gavinsky2007exponential,regev2011quantum}. And yet, despite our advanced understanding of what these quantum advantages are, demonstrating them in a practical setting continues to be an outstanding and central challenge. Important progress has been made in this direction \cite{becerra2013experimental,xiang2011entanglement,ng2012exp,clarke2012experimental,lunghi2013exp,liu2014exp,collins2014realization,berlin2011exp,pappa2014experimental}, but many cases of quantum improvements have never been realized experimentally.

A particular example of a quantum advantage occurs in the field of communication complexity: the study of the minimum amount of information that must be transmitted in order to solve distributed computational tasks \cite{Yao1979,BrassardQCC,RevModPhys.82.665,buhrman1999multiparty,buhrman1998quantum}. It has been proven that for certain problems, quantum mechanics allows exponential reductions in communication compared to the classical case \cite{RevModPhys.82.665,QuantumFingerprinting,RazProblem,HM-Bar-Yossef,gavinsky2007exponential,regev2011quantum}. These results, beside being of great fundamental interest, have important practical applications for the design of communication systems, computer circuits, and data structures \cite{kushilevitz2006communication,RevModPhys.82.665}. However, to date, only a few proof-of-principle implementations of quantum communication complexity protocols have been reported \cite{HornFPs,du2006experimental,TrojekQCC}. Crucially, none of them have demonstrated a reduction in the transmitted information compared to the classical case. Recently, protocols have been introduced that are capable of achieving this reduction using practical optical implementations \cite{arrazolaqfp,arrazola2014QC,arrazola2014TQC}, thus opening the door to experimental demonstrations of the exponential reductions of quantum communication complexity.

Quantum fingerprinting is arguably the most appealing protocol in quantum communication complexity, as it constitutes a natural problem for which quantum mechanics permits an exponential reduction in communication complexity \cite{QuantumFingerprinting,massar2005quantum,arrazolaqfp}. In this problem, Alice and Bob are each given an $n$-bit string, which we label $x$ and $y$ respectively. In the simultaneous message passing model (SMP) \cite{Yao1979}, they must each send a message to a third party, the referee, whose task is to decide whether the inputs $x$ and $y$ are equal or not with an error probability of at most $\epsilon$. Alice and Bob do \emph{not} have access to shared randomness and there is only one-way communication to the referee. In this case, it has been proven that any classical protocol for this problem must transmit at least $\Omega(\sqrt n)$ bits of information to the referee \cite{babai1997randomized,newman1996public}. On the other hand, a quantum protocol was specified in Ref. \cite{QuantumFingerprinting} that transmits only $O(\log_2 n)$ qubits of information -- an exponential improvement over the classical case. However, the proof-of-principle implementations of quantum fingerprinting that have been reported so far only demonstrated the protocol for messages of one single qubit of information, with no reduction in the transmitted information compared to the classical case \cite{HornFPs, du2006experimental}.

In this work, we present an experimental demonstration of a quantum fingerprinting system that is capable of transmitting less information than the best known classical protocol for this problem. Our system is based on the quantum fingerprinting protocol with weak coherent states of Ref. \cite{arrazolaqfp}. Although this protocol is already practical, we overcome various challenges to its experimental implementation. First, we develop an efficient error-correction algorithm that allows us to substantially relax the requirements on the experimental devices and reduce the running time of the protocol. Second, we use an improved decision rule for the referee compared to the one used in Ref. \cite{arrazolaqfp}. Finally, we perform detailed simulations of the protocol that allows us to identify the appropriate parameters for performing the experiment. This enables us to run the protocol using commercial off-the-shelf components.

Indeed, we implemented the protocol by using a commercial plug\&play system originally designed for quantum key distribution (QKD) \cite{idquantique}, to which we added several important modifications. We also characterized the system and showed that, within our theoretical model of the experiment, its performance is consistent with achieving the desired error probability. Finally, we experimentally tested the system for input sizes of up to 100 Mbits, using actual codewords obtained from our error-correction algorithm, but also using repeated random binary sequences to simulate the codewords. We obtained data that are consistent with the protocol transmitting less information than the best known classical protocol.

In the remainder of this paper, we discuss our quantum fingerprinting protocol in detail, focusing on how the choice of parameters affects the error probability and the transmitted information of the protocol. We proceed by describing our error-correction algorithm and conclude by reporting the experimental results as well as discussing the significance of our work.

\subsection*{Coherent-state quantum fingerprinting protocol}\label{sec:protocol}

\begin{figure}
\includegraphics[width=0.95\columnwidth]{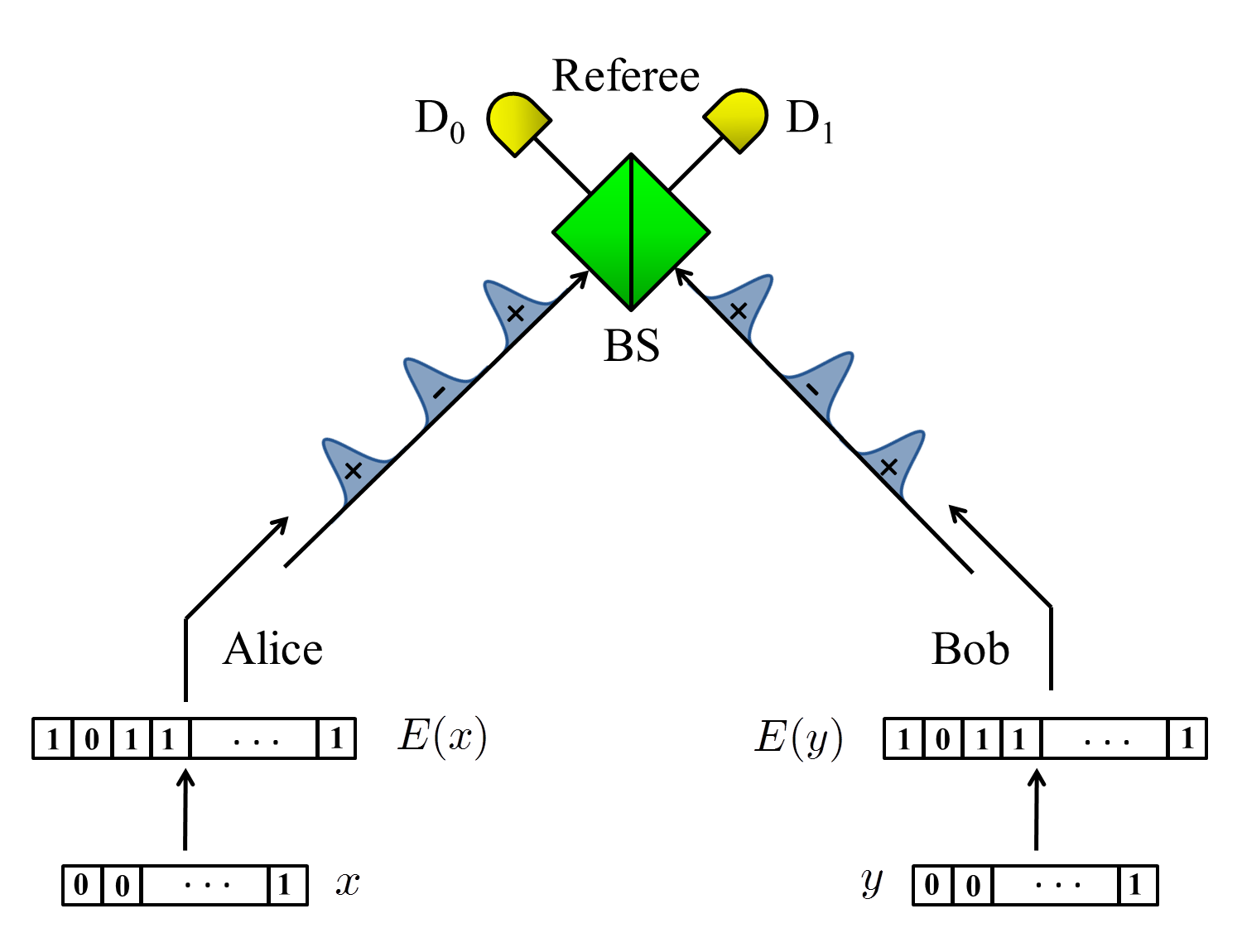}
\caption{(Colour online) A schematic illustration of the quantum fingerprinting protocol. Alice and Bob receive inputs $x$ and $y$, respectively, which they feed to an error-correcting code to produce the codewords $E(x)$ and $E(y)$. Using these codewords, they modulate the phases of a sequence of coherent pulses that they send to the referee. The incoming signals interfere at a beam-splitter (BS) and photons are detected in the output using single-photon detectors $D_0$ and $D_1$. In an ideal implementation, detector $D_1$ fires only when the inputs to Alice and Bob are different. }\label{Fig:diagram}
\end{figure}

In the quantum fingerprinting protocol of Ref. \cite{arrazolaqfp}, portrayed in Fig.~\ref{Fig:diagram}, Alice first applies an error-correcting code $E: \{0,1\}^n\rightarrow \{0,1\}^m$ to her input $x$ of $n$ bits. This results in a codeword $E(x)$ of $m=\frac{n}{R}$ bits, which she uses to prepare a sequence of $m$ coherent states, where $R=\frac{n}{m}<1$ is the rate of the code. This sequence of coherent states is given by the state

\beq\label{coherentfpstates}
\ket{\alpha,x}=\bigotimes_{i=1}^m\ket{(-1)^{E(x)_i}\frac{\alpha}{\sqrt{m}}}_i.
\eeq
Here $E(x)_i$ is the $i$th bit of the codeword and $\alpha$ is a complex amplitude. Notice that all the coherent states have the same amplitude, but their individual phases depend on the particular codeword, which in turn is determined by the input $x$. The total mean photon number in the entire sequence is $\mu:=|\alpha|^2$, which in general depends on the length of the codewords $m$.

Bob does the same as Alice for his input $y$, and they both send their sequence of states to the referee, who interferes the individual states in a balanced beam-splitter. The referee checks for clicks at the outputs of the interferometer using single-photon detectors, which we label ``$D_0$" and ``$D_1$". In the ideal case, a click in detector $D_1$ will never happen if the phases of the incoming states are equal, i.e. if $E(x)_i\oplus E(y)_i=0$. However, it is possible for a click in detector $D_1$ to occur if the phases are different, i.e. if $E(x)_i\oplus E(y)_i=1$. Thus, if $x\neq y$, we expect a number of clicks in $D_1$ that is proportional to the total mean number of photons and the Hamming distance between the codewords. This allows the referee to distinguish between equal and different inputs by simply checking for clicks in detector $D_1$.

In Ref. \cite{arrazolaqfp}, it was proven that the quantum information $Q$ that can be transmitted by sending the states of Eq. \eqref{coherentfpstates} satisfies
\beq\label{scaling}
Q=O(\mu \log_2 n),
\eeq
which, for fixed $\mu$, is an exponential improvement over the classical case. Essentially, by fixing the total mean photon number to a constant, we are restricting ourselves to an exponentially small subspace of the larger Hilbert space associated with the optical modes, which in turn restricts the capability of these systems to transmit information.

In the presence of experimental imperfections such as detector dark counts and optical misalignment, detector $D_1$ may fire even when the inputs are equal. Therefore, it does not suffice to check for clicks in this detector -- we must introduce a different decision rule for the referee. The decision rule proposed in Ref. \cite{arrazolaqfp}, which is based on the fraction of clicks that occur in detector $D_1$, is extremely sensitive to experimental imperfections. Instead, in this work we construct a better decision threshold based only on the total number of clicks observed in detector $D_1$.

Let $D_{1,E}$ and $D_{1,D}$ be random variables corresponding to the number of clicks in detector $D_1$ for the case of equal and worst-case different inputs, respectively. It can be shown that these distributions can be well approximated by binomial distributions $D_{1,E}\sim \textrm{Bin}(m,p_E)$ and $D_{1,D}\sim\textrm{Bin}(m,p_D)$, where $m$ is the number of modes and $p_E$, $p_D$ are the probabilities of observing a click in each mode for the case of equal and worst-case inputs respectively. These probabilities are given by \cite{arrazolaqfp}:
\begin{align}
p_E=& (1-e^{-\frac{2(1-\nu)\mu}{m}})+p_{dark}\label{pE}\\
p_D=& \delta(1-e^{-\frac{2\nu\mu}{m}})+(1-\delta)(1-e^{-\frac{2(1-\nu)\mu}{m}})+p_{dark}\label{pD}.
\end{align}
Here $\nu$ is the interference visibility -- which quantifies the contrast of the interferometer -- and $p_{dark}$, the dark count probability, is the probability that a detector will fire even when no incident photons from the signals are present. As before, $\mu$ is the total mean photon number in the signals and $\delta$ is the minimum distance of the error-correcting code, which is defined as the smallest relative Hamming distance between any two distinct codewords.

\begin{figure}
\includegraphics[width=\columnwidth]{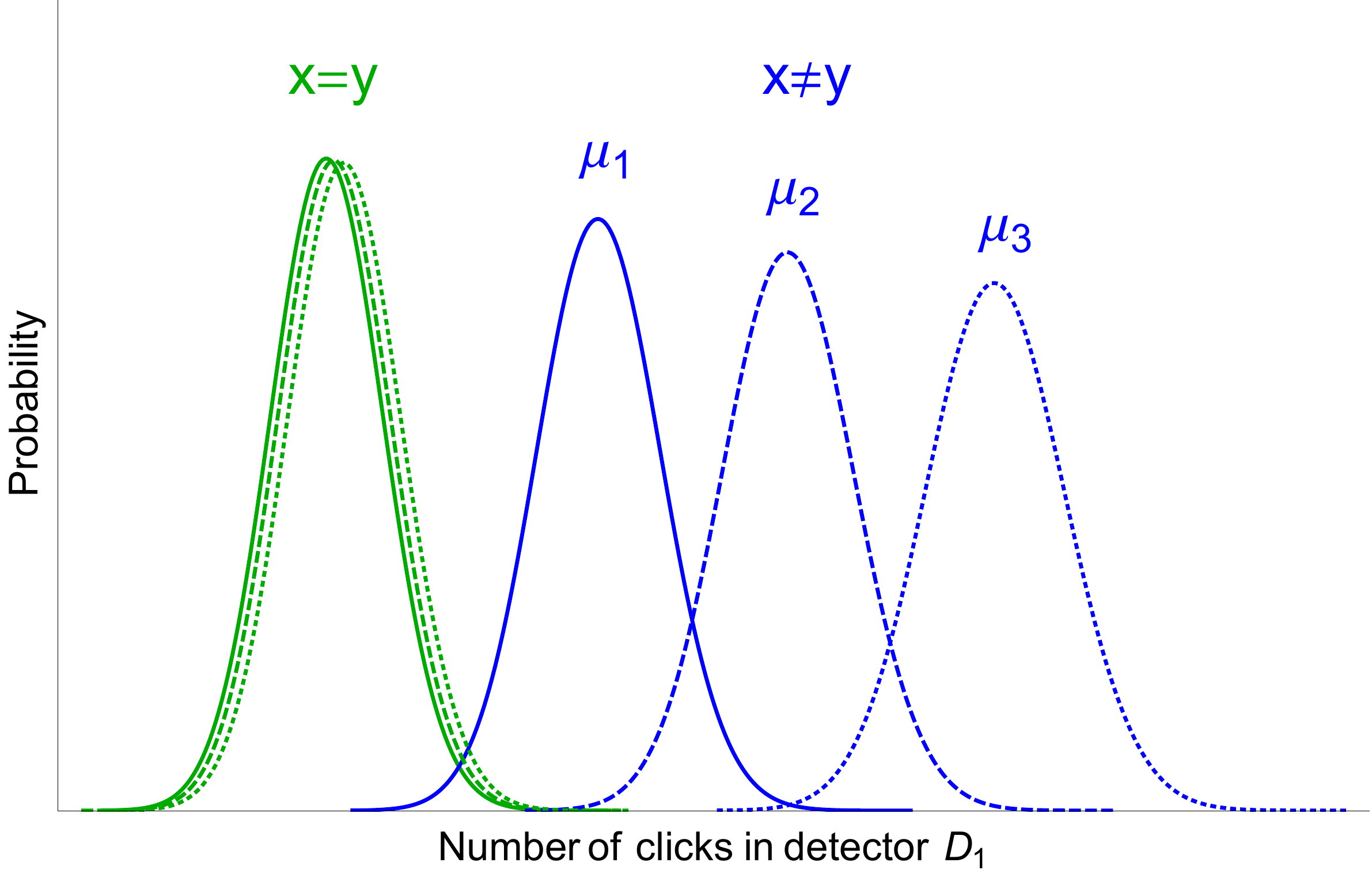}
\caption{(Colour online) An illustration of the probability distributions for the number of clicks in detector $D_1$ for equal inputs ($x=y$) and worst-case different inputs ($x\neq y$). The distributions are shown for three different total mean photons numbers: $\mu_1$ (solid), $\mu_2$ (dashed) and $\mu_3$ (dotted), with $\mu_1<\mu_2<\mu_3$. The distributions for equal inputs (green) are dominated by dark counts, so they are largely unaffected by changes in $\mu$. On the other hand,  for the worst-case different inputs (blue), the mean value of the distributions depends strongly on $\mu$. Therefore, the error in distinguishing both distributions can be controlled by choosing $\mu$ appropriately. }\label{Distributions}
\end{figure}

The referee sets a threshold value $D_{1,th}$ such that, if the number of clicks is smaller or equal than $D_{1,th}$, he will conclude that the inputs are equal. Otherwise, he concludes that they are different. Note that, unlike the ideal case, in the presence of imperfections, an error can occur even when the inputs are equal. In our protocol, the value of $D_{1,th}$ is chosen in such a way that an error is equally likely to occur in both cases, so that the probability of error is given by
\beq\label{Eq: Perror}
\Pr(\textrm{error})=\Pr(D_{1,E}>D_{1,th})=\Pr(D_{1,D}\leq D_{1,th}),
\eeq
which can be calculated directly from the distributions of $D_{1,E}$ and $D_{1,D}$. This is illustrated in Fig. \ref{Distributions}. Finally, note that this model is expected to be correct as long as the parameters quantifying the experimental imperfections as well as the mean photon number $\mu$ are all constant during the run of the protocol. In practice this is not necessarily the case, so our model should be understood as an approximation of the actual performance of the system.

Finally, we note that in any implementation of the protocol there will be some loss captured by the combined effect of limited detector efficiency and channel loss. We quantify this with the single parameter $\eta<1$. As shown in Ref. \cite{arrazolaqfp}, the effect of loss can be compensated by adjusting the total mean photon number accordingly: $\mu\rightarrow \mu/\eta$. Thus, the protocol is robust to loss.

\subsection*{Error-Correcting Code}
In quantum fingerprinting, an error-correcting code (ECC) is used to amplify the Hamming distance between the inputs of Alice and Bob. Even if these inputs are originally very close to each other -- for example if they differ in a single position -- after applying the ECC, the resulting codewords will have a much larger Hamming distance. In the worst-case scenario, this distance is given by the minimum distance of the code. Note, however, that no error-correction actually takes place in the quantum fingerprinting protocol -- we just use the properties of error-correcting codes to increase the distance between the inputs.

The quantum fingerprinting protocol of Ref. \cite{arrazolaqfp} used Justesen codes as an example to illustrate the properties of the protocol. However, these codes are not optimal for quantum fingerprinting. In this section, we construct more efficient codes based on random Toeplitz matrices that significantly relax the requirements on the experimental devices and lead to a faster implementation of the protocol. Due to their probabilistic construction, these codes are not guaranteed to have the desired minimum-distance, but do achieve it with exponentially high probability (See Supplementary Material). Therefore, in our protocol, we only claim that, with exponentially high probability, we are using codes with the desired properties.

\begin{figure}
\includegraphics[width=\columnwidth]{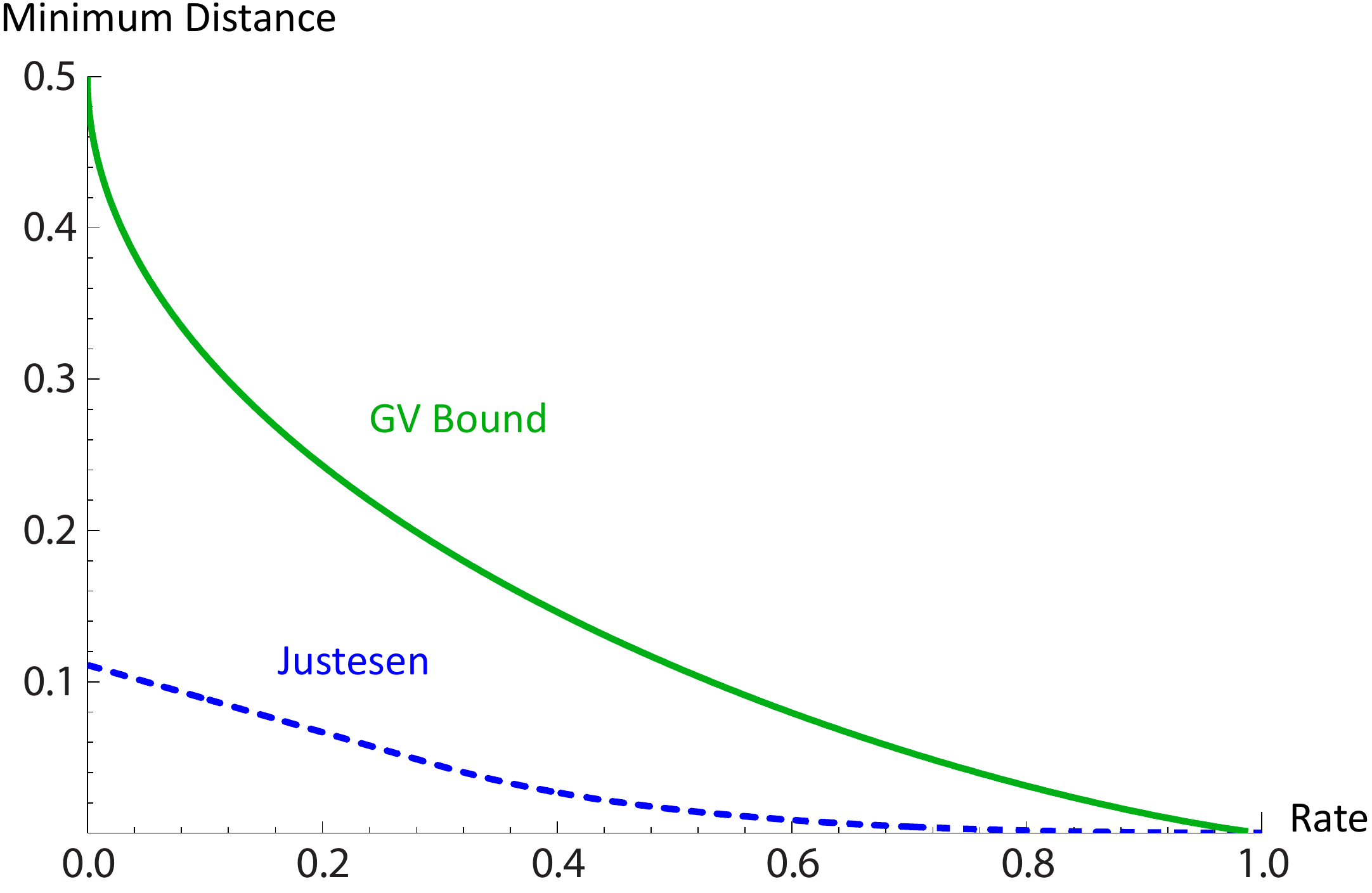}
\caption{(Colour online) The Gilbert-Varshamov bound compared to the distance-rate relationship achieved by Justesen codes. For various rates, a code satisfying the GV bound -- like the one we use in our protocol -- achieves a minimum distance that is more than three times the value for Justesen codes.}\label{Fig: GV bound}
\end{figure}

An ECC with a high rate and a large minimum distance is desired, since a higher rates leads to lower transmitted information and larger tolerance for dark counts, while a larger minimum distance leads to smaller error probability for fixed mean photon number. Fundamentally, there is an inherent trade-off between the rate and distance of ECCs. In particular, the Gilbert-Varshamov (GV) bound \cite{Gilbert, Varshamov} states that there exists some binary linear code whose rate $R$ and minimum distance $\delta$ satisfy the relation
\beq
R \geq 1-H_2(\delta),
\eeq
where $H_2(\cdot)$ is the binary entropy function. Using a binary linear code that approaches this bound would constitute a significant improvement over the codes used in previous protocols. This is clearly illustrated in Fig. \ref{Fig: GV bound}.

It is well known in coding theory that random linear codes (RLCs) can asymptotically approach the GV bound with encoding complexity $O(n^2)$ \cite{barg2002random}. However, in quantum fingerprinting, the input size $n$ is typically very large (e.g. $n = 10^8$), thus making the encoding time prohibitively high. In order to reduce this encoding complexity, we make use of a subclass of RLCs whose generator matrices are Toeplitz matrices. A Toeplitz matrix is a matrix in which each descending diagonal from left to right is constant. An $n \times m$ Toeplitz matrix is completely determined by the $n+m-1$ elements on its first row and column. This structure implies that only $O(n \log n)$ time for encoding is required for this subclass of RLCs \cite{handbook}. Additionally, these codes also asymptotically approach the GV bound (see Supplementary Material for a proof). By using this family of codes, we are able to reduce the encoding times by several orders of magnitude, making them suitable for practical applications.

The exponential separation between quantum and classical communication complexity for the equality function only holds if Alice and Bob do not have access to shared randomness that is generated in each run of the protocol \cite{babai1997randomized}. However, even though the generator matrices of our RLCs are randomly constructed, once they have been created they remain fixed for all future instances of the protocol. This ensures that no new randomness is generated in each run of the protocol, as required to satisfy the conditions of the exponential separation. In particular, Alice and Bob can store the generator matrices in memory and use them to encode their inputs in exactly the same way as if they had been generated deterministically.

For our experiment, an encoder program written in C++ was built and tested, demonstrating the feasibility of this subclass of RLCs. The free Fast-Fourier Transform library FFTW was used to accelerate multiplications with Toeplitz matrices \cite{frigo2005design} and the random numbers to construct the matrices were generated from a quantum random number generator \cite{xu2012ultrafast}. The results from an optimized encoder are shown in Table \ref{Tab:ECC}. As we can see, our encoder is highly practical, can be run on any common lab PC, and finishes the encoding in an acceptable time frame for input sizes as large as $n= 3 \times 10^8$. Faster encoding times could be obtained by using dedicated hardware.

\begin{table}[h!]
\begin{tabular}{c|c|c|c}\hline\hline
$n$ (bit)& $m$ (bit)& Time (s)& Memory (Mbit) \\ \hline
$10^6$ & $5\times 10^6$& 6 & 52 \\
$10^7$ & $5\times 10^7$& 106 & 733 \\
$3\times 10^7$ & $1.5\times 10^8$& 181 & 1654 \\
$3\times 10^8$ & $1.5\times 10^9$& 4831 & 10000 \\ \hline\hline
\end{tabular}
\caption{The performance of the encoder for different input sizes, using a computer with a quad-core i7-4770 @3.4GHz CPU and 16GB RAM. Running times are acceptable for experimental applications for input sizes as large as $n=3 \times 10^8$.}\label{Tab:ECC}
\end{table}

\subsection*{Experimental setup} \nonumber
We demonstrate our proof of concept quantum fingerprinting protocol using a plug\&play scheme~\cite{stucki02a}, initially designed for quantum key distribution (QKD). The advantage of the plug\&play system with respect to other viable systems is that it offers a particularly robust and stable implementation. This allows us to perform reliable experiments with highly attenuated coherent states for long time durations. We implement the protocol on top of two commercial systems, namely ID-500 and Clavis2, manufactured by ID Quantique~\cite{idquantique}.

\begin{figure*}[!t]
\centering
\resizebox{15cm}{!}{\includegraphics{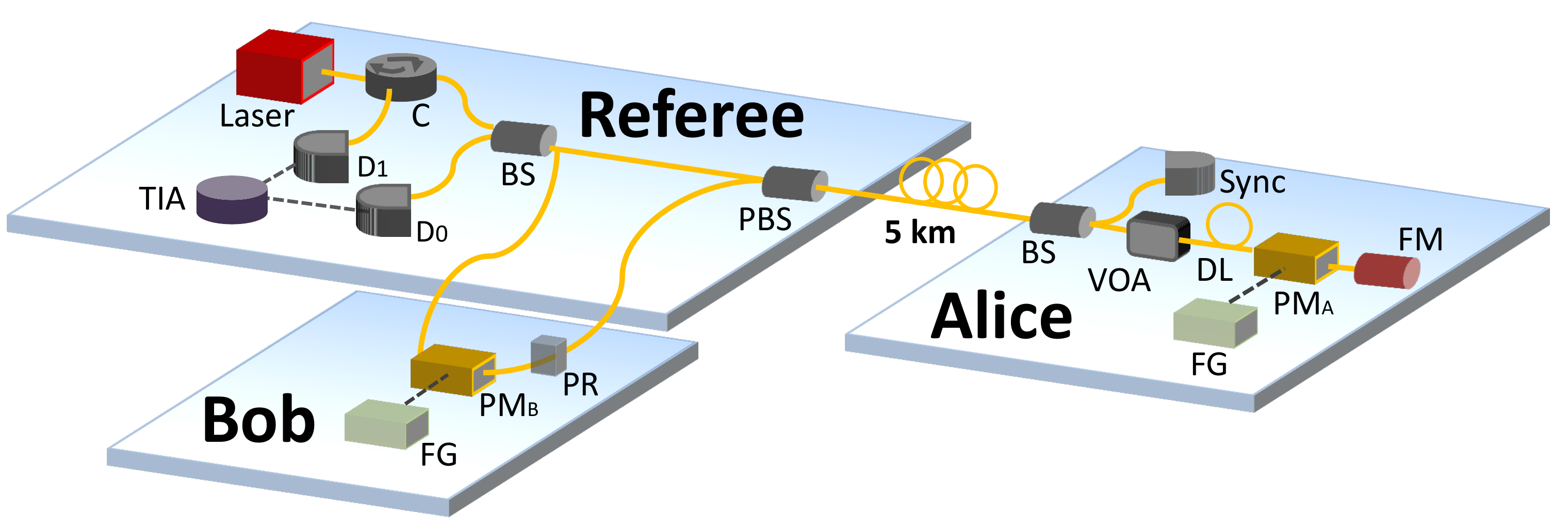}} \caption{(Color online) Experimental setup for quantum fingerprinting. The laser source at the referee's setup emits photon pulses which are separated at a 50:50 beam-splitter (BS) into two pulses, the signal pulse and the reference pulse. The reference pulse passes through Bob's phase modulator (PM) and then through a polarization rotator which rotates the pulses' polarization by $90\degree$. The pulses are then recombined at a polarization beam splitter (PBS) where they exit through the same port and travel to Alice through the 5 kms fiber. Alice uses the reference pulse as a synchronization (Sync) and uses her phase modulator (PM) to set the phase of the signal pulse according to her codeword $E(x)$. Once the two pulses are reflected back by the Faraday mirror (FM), she attenuates them to the desired photon level by using the variable optical attenuator (VOA). When the two pulses return in the direction of the referee, because of Alice's FM, the reference pulse will travel through Bob, who uses his PM to modulate the pulse according to his codeword $E(y)$. Both Alice and Bob use two external function generators (FG) to control the PMs. Finally, the two pulses arrive simultaneously at the BS, where they interfere and are detected by two detectors $D_0$ and $D_1$. The detection events are recorded by a time interval analyzer (TIA).} \label{Fig:scheme}
\end{figure*}

In our set-up, which is shown in Fig.~\ref{Fig:scheme}, the referee starts by sending two strong pulses at about 1551 nm to Alice over a 5 km fiber. The two pulses are created from a single pulse by a beam splitter (BS), and they take paths of different length before being recombined by a polarization beam splitter (PBS). The linear polarization of the pulse travelling through the longer path, which corresponds to Bob's system, is rotated by $90\degree$. After reaching the PBS, the two pulses exit through the same port and travel to Alice with a time delay between them caused by the difference in path lengths. The front and back pulses are called the reference pulse and the signal pulse, respectively.

Once they reach Alice, she uses the reference pulse as a synchronization signal to activate her phase modulator (PM), which she employs to set the phase of the signal pulse according to her codeword $E(x)$. Both pulses are reflected back by a Faraday mirror (FM), which rotates the pulses' polarization by $90\degree$, and she attenuates them to the desired photon level using the variable optical attenuator (VOA). Once the pulses return back, due to the FM, the pulses take opposite paths, such that the reference pulse now passes through Bob and its phase is modulated by Bob's PM according to $E(y)$. Finally, the two pulses interfere at the referee's BS and the detection events are registered using two high-quality single-photon detectors $D_0$ and $D_1$. It is important to note that the returning signal pulse modulated by Alice travels directly to the referee, while the returning reference pulse passing through Bob does not contain any information about Alice's codeword. This guarantees that there is \emph{no} communication between Alice and Bob.

Since the operating conditions of our protocol are significantly different from those of standard QKD, using a commercial QKD equipment for our implementation requires several important modifications to the system. First, two single-photon detectors with low dark count rates were installed. Indeed, as can be deduced from Eqs. \eqref{pE} and \eqref{pD}, lower dark count rates permit the operation of the system at lower mean photon numbers, which lead to a reduction in the transmitted information. Fortunately, our error correction codes improve the tolerance of the protocol to dark counts, which permits us to use commercial detectors. We employ two commercial free-running InGaAs avalanche photodiodes -- ID220~\cite{idquantique}. The dark count rate per 1 ns detection gate is about $(3.5\pm0.2)\times 10^{-6}$ and the corresponding quantum efficiency is about 20\%. The detections are recorded by a high-precision time interval analyzer (TIA, PicoQuant HydraHarp 400). The system was run at a repetition rate of 5 MHz with the detector dead time set at 10$\mu$s. This means that after a click occurred, the following 50 pulses are blocked before the detector is active again. This is not a problem in our experiment because the mean photon number in each pulse is extremely low, therefore the expected number of undetected photons as a result of this effect is negligible compared to other sources of error (see Supplementary Material for details).

Additionally, new functionalities and control signals were added to the system. On one hand, we use the VOA inside Alice to reduce the mean photon number per pulse down to suitable numbers. These values -- in the order of $10^{-5}$ per pulse -- were in fact four orders of magnitude lower than those typically used for QKD. Hence, several calibration processes of the system are required, which imposes particular care in the synchronization of the phase modulation and attenuation signals. On the other hand, commercial QKD systems like Clavis2 have an internal random number generator to set the phase modulations, which does not allow us to modulate the phases according to the pre-generated codewords. We solve this difficulty by using two external function generators (FG, Agilent 88250A) loaded with the codewords to control Alice's and Bob's PM. This requires precise synchronization and calibration procedures to guarantee correct phase modulations. We finally observed high interference visibility of about $(99\pm0.5)\%$ after careful calibration.

In the implementation on ID-500, the random numbers controlling the phase modulations are accessible to users. We use our codewords to replace those random numbers directly. However, after testing for an input data size of $n=1.42\times 10^{8}$ on ID-500, an unexpected hardware problem made ID-500 unavailable for further experiments. To further test the feasibility of our protocol for different input sizes, we switched to Clavis2 for measurements. In the implementation on Clavis2, since each function generator has a small memory, for simplicity we load a frame of about 430 random numbers to each function generator and reuse these random numbers. This allows us to create binary sequences with the desired distance $\delta$ that can be used to test the performance of the system. All the above modifications led to the development of a practical system that is capable of performing quantum fingerprinting.

\subsection*{Experimental results} \nonumber
\begin{table*}\center
\begin{tabular}{c @{\hspace{0.3cm}} c @{\hspace{0.3cm}} c  @{\hspace{0.3cm}} c @{\hspace{0.3cm}} c @{\hspace{0.3cm}} c}
\hline \hline
$\eta_{AR}$ & $\eta_{BR}$ & $\eta_{det}$ & $p_{dark}$ & $\nu$ \\
\hline 3 dB (2.36 dB) & 1.5 dB (1 dB) & $20.0\%$ & $(3.5\pm0.2)\times 10^{-6}$ & $(99\pm0.5)\%$ \\
\hline \hline
\end{tabular}
\caption{Parameters measured in the implementations. The overall loss between the output of Alice's VOA and the input to the referee's detectors is given by the parameter $\eta_{AR}$. Similarly, $\eta_{BR}$ defines the overall loss between the output of Bob's PM and the referee's detectors. Both $\eta_{AR}$ and $\eta_{BR}$ are carefully characterized in ID-500 (Clavis2). The other parameters are the detector's quantum efficiency $\eta_{det}$, dark count rate per pulse $p_{dark}$ for each detector, and system visibility $\nu$, which are nearly the same for ID-500 and Clavis2. }\label{Tab:parameters}
\end{table*}

We perform the quantum fingerprinting experiment over a standard telecom fiber of 5 km between Alice and the referee. The overall loss between the output of Alice's VOA and the input of the referee's detector $D_1$ -- which includes the losses of quantum channel, PBS, BS and the circulator -- is about 3 dB (2.36 dB) for ID-500 (Clavis2). The channel between Bob and the referee is about a few meters, and its overall loss including Bob's channel, the BS and the circulator, is about 1.5 dB (1 dB). We summarize all system parameters in Table~\ref{Tab:parameters}. Based on these parameters, for a given input size $n$, we use our model of the protocol to optimize the photon number $\mu$ in order to achieve a desired error probability $\epsilon$.

Because there is loss in the channels and the detectors are not perfectly efficient, Alice and Bob must use higher mean photon numbers compared to the case with no channel loss and perfect detectors. As implied by Eq. \eqref{scaling}, this also leads to an increase in the transmitted information, which we take into account in our calculations of the transmitted information. In particular, if Alice and Bob experience different amounts of loss, they must choose a different mean photon number when preparing their signals, ensuring that the amplitude of their pulses is equal when they interfere in the referee's beam splitter.

In the experiment, the detection events registered on $D_0$ and $D_1$ in conjunction with the known experimental conditions in the system can be used to characterize the photon numbers sent out by Alice and Bob, the dark count probability, and the visibility of the interferometer. From the characterization of these parameters, we find that there is a good agreement with our model of the system. The main source of uncertainty is due to an imperfect matching between the observed mean photon numbers and those pre-calibrated from the VOA. This uncertainty is determined by the fluctuations of several devices, such as laser power, VOA, and detector efficiency. The detailed values of this uncertainty are shown in the Supplementary Material.

The quantum fingerprinting protocol is tested over several values of the input size $n$. For each $n$, we record the detection counts on $D_1$ for two types of input data: equal inputs $E(x)=E(y)$, and the worst-case different inputs, i.e. those for which the codewords $E(x)\neq E(y)$ have a distance equal to the minimum distance. For our experiment, we minimize the transmitted information by choosing an optimal value of $\delta=0.22$ for the minimum distance. From the threshold value $D_{1, th}$ that is pre-calculated from our model, the referee can distinguish between equal and different inputs. The upper bound $Q$ on the quantum information Alice and Bob is calculated from their respective mean photon numbers $\mu_A$ and $\mu_B$, as well as the codeword length $m$.

\begin{figure}[!t]
\centering
\resizebox{8.5cm}{!}{\includegraphics{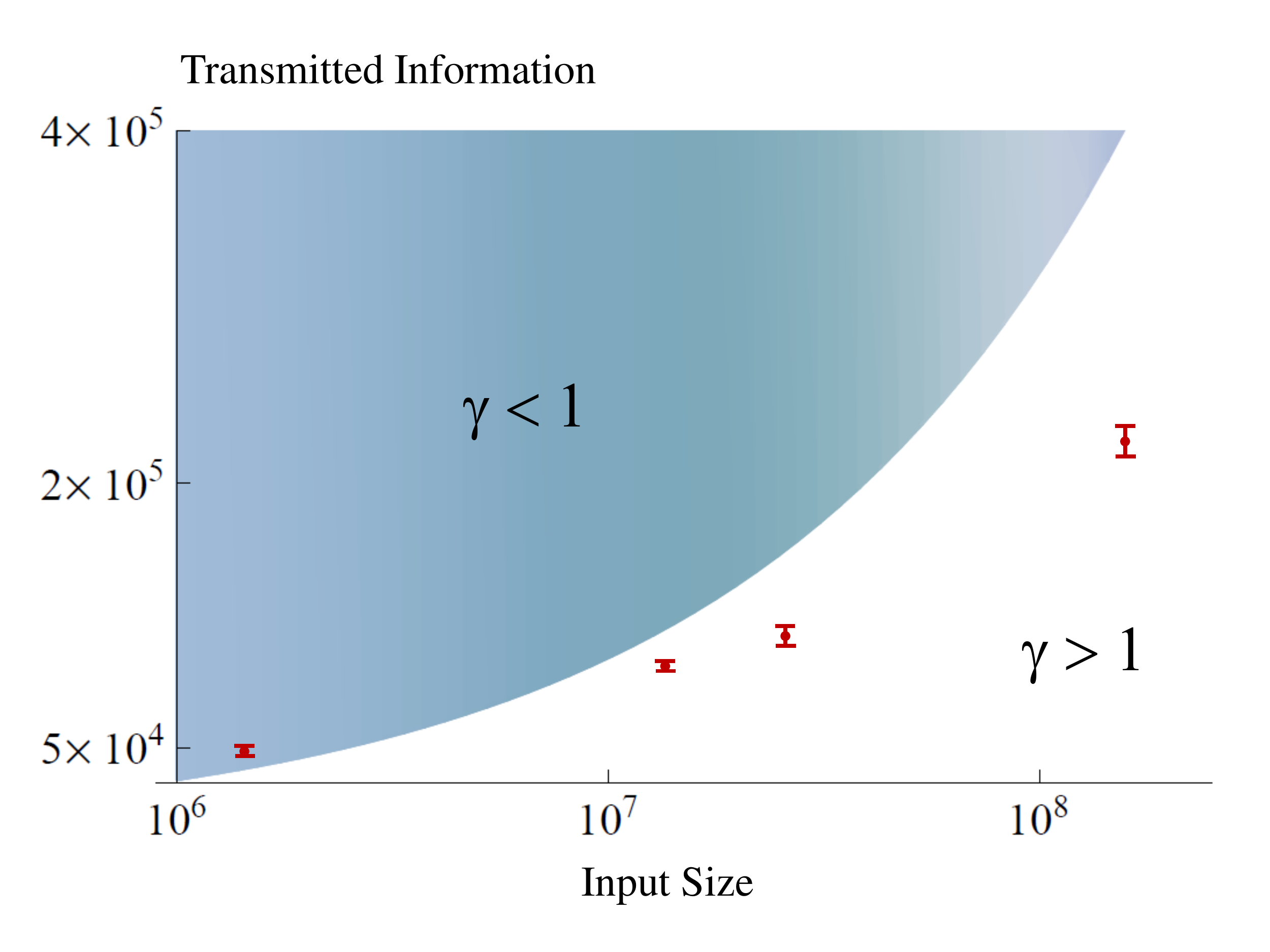}} \caption{(Color online) Transmitted information in our protocol. The blue area indicates the region where the classical protocol transmits less information than our protocol, while the red point shows our experimental results. The error bars correspond to one standard deviation. For large $n$, our results are strictly better than the best known classical protocol for a range of practical values of the input size.} \label{Fig:transmittedinformation}
\end{figure}

In Fig.~\ref{Fig:transmittedinformation}, we show the transmitted information as a function of the input size $n$ for a target error probability of $\epsilon=5\times10^{-5}$. The error probability was calculated from our theoretical model of the experiment. within experimental uncertainty, the worst-case values of the mean photon number, visibility, and dark count probability were used to reconstruct the probability distributions of clicks in detector $D_1$. These distributions, in turn, were used to calculate the error probability from Eq. \eqref{Eq: Perror}. Since our theoretical model is only an approximation, the error probability should also be understood as approximate. The blue area indicates the information transmitted by the best known classical protocol of Ref. \cite{babai1997randomized} which for this probability of error requires the transmission of $16\sqrt{n}$ bits. The red points show our experimental results, where the data point for the largest $n$ is obtained from ID-500 and the other three data points are obtained from Clavis2. Note that Clavis2 and ID-500 have almost the same optics and functionality~\cite{idquantique}. We use the same measurement and processing method for the data obtained from these two systems, and show the experimental results together in one figure instead of two. The error bars come from the uncertainty in the estimation of the mean photon number $\mu$. For large $n$, our experimental results are strictly better than those of the classical protocol for a wide range of practical values of the input size.

\begin{figure}[!hbt]
\centering
\resizebox{8.5cm}{!}{\includegraphics{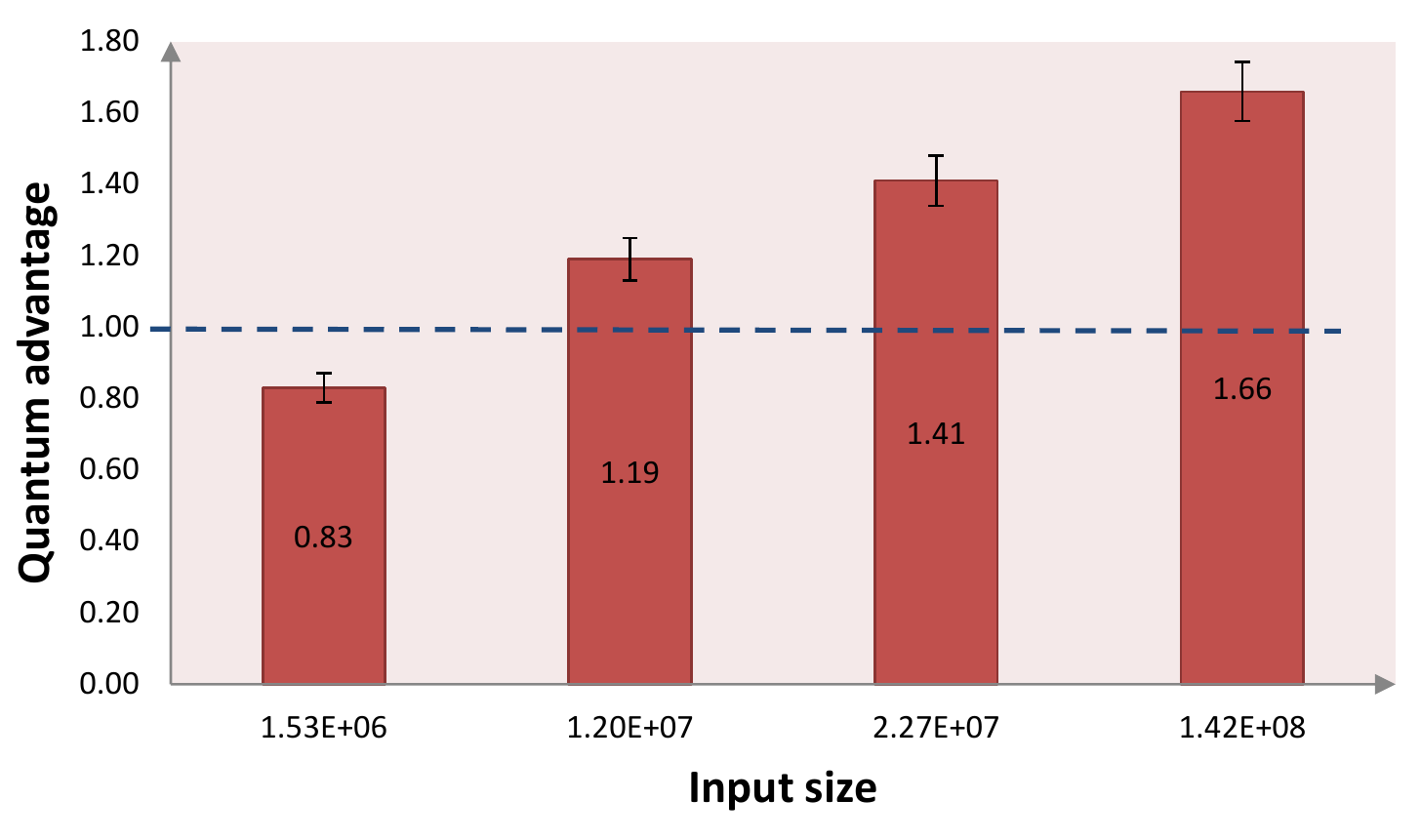}} \caption{(Color online) The quantum advantage $\gamma$ between the transmitted classical information and the upper bound on the transmitted quantum information. For the three large input sizes, the ratio is well above 1. The quantum advantage was as large as $\gamma =1.66$, which implies that the transmitted information in the classical protocol was 66\% larger than in the quantum case. } \label{Fig:ratio}
\end{figure}

To obtain further insight into our results, we define the \emph{quantum advantage} $\gamma$ as the ratio between the transmitted classical information $C$ of the best-known classical protocol \cite{babai1997randomized} and the upper bound $Q$ on the transmitted quantum information:
\begin{eqnarray}
\gamma=\frac{C}{Q}.
\end{eqnarray}
A value $\gamma>1$ for a given error probability $\epsilon$ implies that less information is transmitted in the quantum case than in the classical one. This allows us to use the quantum advantage as a figure of merit to assess the performance of our quantum fingerprinting implementation. In Fig.~\ref{Fig:ratio}, we show the experimental results for $\gamma$ as a function of different input sizes. For the three largest input sizes, the ratio is well above 1, and the classical protocol transmitted as much as 66\% more information than the quantum protocol.  For the smallest input size, no quantum improvement was obtained.

\subsection*{Discussion}

Based on the protocol of Ref. \cite{arrazolaqfp}, we have experimentally demonstrated a proof of concept quantum fingerprinting system that is capable of transmitting less information than the best known classical protocol for this problem. Our experimental test of this system indicates that its operation is consistent with our model of the devices and hence also with achieving the desired error probability. Moreover, we have operated our system in a parameter regime in which the information transmitted in the protocol is up to $66\%$ lower than the best known classical protocol. This constitutes the first time that a quantum fingerprinting protocol has been carried out that is capable of achieving this reduction in the transmitted information.

In communication complexity, it is assumed that the parties have unlimited computational power. However, from a practical perspective, it may not always be possible to ignore these computational requirements. In fact, even though the running time during communication of our experiment scales linearly with the input size, the total running time of the protocol is dominated by the time required to run the error-correcting code -- which is a crucial component of the protocol. For instance, at a repetition rate of 5MHz, it takes 300 seconds to run the communication for an output size of $m=1.5\times 10^9$. On the other hand, even with the use of RLCs with quasi-linear encoding complexity, 4831 seconds are needed to run the encoding algorithm, as seen in Table \ref{Tab:ECC}. Therefore, the practical advantages of quantum fingerprinting, in terms of reductions in resource expenditures, will likely be found in a reduction of the number of photons used. This is a major property that our protocol possesses. Indeed, for the largest input size that we tested, $n=1.42\times 10^8$, a mean photon number of $\mu\approx7\times 10^3$ was used (see Supplementary material for details). Overall, it is remarkable that quantum fingerprinting with coherent states can be realized while revealing only a very small amount of information to the referee -- a feature of the protocol that may have important applications to fields such as cryptography \cite{gavinsky2013secrets} and information complexity \cite{chakrabarti2001informational}, where this extremely small leakage of information plays a fundamental role.

In our quantum fingerprinting protocol, the maximum reduction in the transmitted information depends crucially on the dark count probability and the overall loss in the system. Thus, our results can be directly improved by using detectors with higher efficiency and lower dark counts. This can lead to a quantum fingerprinting protocol that, with the use of available technology \cite{marsili2013detecting}, transmits several orders of magnitudes less information than the best known classical protocol for large input sizes. Even though there is no proof that the best known classical protocol is optimal, a lower bound for the classical transmitted information was proven in Ref. \cite{babai1997randomized}. This lower bound states that, for any classical protocol with error probability smaller than 0.01, Alice and Bob must send at least $\frac{\sqrt{n}}{20}$ bits of information. This is roughly two orders of magnitude smaller than the transmitted information of the best known classical protocol. By using state-of-the-art detectors, it should be possible to demonstrate a quantum fingerprinting protocol capable of beating this classical lower bound. Achieving this would constitute a significant milestone for experimental quantum communication complexity.

The main challenges to achieving such a large reduction in the transmitted information are the computational requirements of the error-correcting algorithms as well as using single-photon detectors with very small dark count rates and high efficiency. Additionally, it is desirable to carry out the protocol in a configuration that, unlike the plug\&play scheme, permits Bob to be situated at a large distance from the referee. Finally, in this work, we have tested our model of the system and used that test to make an indirect assessment of the error probability based on our theoretical model. Future implementations should improve on this by treating the system as a black-box, using the data directly to make statistical inferences about the error probability, without relying on an approximate model of the system. Overall, our results constitute a significant first step in the development of experimental quantum communication complexity, which may also be extended to other protocols with a proven exponential advantage over the classical case \cite{RazProblem,HM-Bar-Yossef,arrazola2014QC,arrazola2014TQC}.

\subsection*{Acknowledgements}
This work was supported in part by CFI, OIT, NSERC RTI, the NSERC Strategic Project Grant (SPG)
FREQUENCY, the NSERC Discovery Program, the CRC program, Connaught Innovation fund and Industry
Canada. The authors thank A. Ignjatovic, Pedro Palacios-Avila, Z. Tang, Francisco de Zela for valuable discussions. Particularly, F. Xu and K. Wei thank V. Makarov for his hospitality during their visit to IQC. The authors acknowledge support from the Mike and Ophelia Lazaridis Fellowship, OGS VISA award, the Office of Naval Research (ONR), the Air Force Office of Scientific Research (AFOSR), the National Natural Science Foundation of China (Grant No. 61178010), China Scholarship Council (No. 201406470051), the University of Hong Kong Physics Department Summer Overseas Research Program and the Science Faculty Overseas Research Fellowship, CONCYTEC-Peru, USEQIP, the IQC Summer URA, the NSERC Postdoctoral Fellowships Program and CryptoWorks21.

\section*{Supplementary Material}
\subsection*{Error Probability Analysis}

Let $G$ be a random $n \times m$ Toeplitz matrix over $\mathbb{F}_2$.
There are two failure events associated with $G$: the minimum distance $\delta$ being not as large as promised (which results in less-than-expected worst case performance) and the matrix $G$ being not full rank (which can cause two different inputs to be mapped to the same output, leading to a minimum distance of $\delta=0$). We will show that,
for any fixed rate $R$ less than $1 - H_2(\delta)$,
the probabilities of both failure events
decreases exponentially with the output size $m$
and can thus be neglected for sufficiently large $m$. \\

\begin{thm}\label{thm:GV}\cite{essentialcoding}
Let $G \in \mathbb{F}_2^{n \times m}$ be a Toeplitz matrix
chosen uniformly at random. Let $\delta_{\min}(G)$ be the minimum
 distance of the linear code with $G$ as generator matrix.
Then, for any $\delta \in (0, 1/2)$,
\[
\Pr (\delta_{\min}(G) \le \delta) \le 2^{-m\left(1 - H_2(\delta) - R \right)}.
\]
In particular, if $R = 1 - H_2(\delta) - \epsilon$, for some
$\epsilon > 0$, then
\[
\Pr (\delta_{\min}(G) \le \delta) \le 2^{-
\epsilon m}.
\]
\end{thm}

\begin{table*}[hbt!]\center
\begin{tabular}{c @{\hspace{0.5cm}} c @{\hspace{0.5cm}} c  @{\hspace{0.5cm}} c @{\hspace{0.5cm}} c}
\hline \hline
$n$ & $1.53\times 10^{6}$ & $1.20\times 10^{7}$ & $2.27\times 10^{7}$ & $1.42\times 10^{8}$ \\
\hline
$\mu_{A}$ & 1914$\pm$68 & 3295$\pm$118 & 3670$\pm$131 & 7120$\pm$254  \\
\hline
$D_{1,E}$ & 22 & 277& 830 & 1939 \\
\hline
$D_{1,D}$ & 131 & 318 & 954 & 2224 \\
\hline
$D_{1,th}$ & 49 & 302 & 902 & 2110 \\
\hline
$Q$ & 47689$\pm$1703 & 93152$\pm$3326 & 108129$\pm$3860 & 229713$\pm$8201 \\
\hline
$\gamma$ & 0.83$\pm$0.02 & 1.19$\pm$0.05 & 1.41$\pm$0.05 & 1.66$\pm$0.06  \\
\hline
$\epsilon$ & $(1.6\pm0.9) \times 10^{-9}$ & $(2.3\pm1.4) \times 10^{-7}$ & $(6.6\pm3.7) \times 10^{-6}$ & $(2.9\pm1.3) \times 10^{-5}$ \\
\hline \hline
\end{tabular}
\caption{Detailed experimental results. The parameter $\mu_A$ is the mean photon number used by Alice. For the clicks in detector $D_1$ we report the observed averages for the case of equal inputs $D_{1,E}$, different inputs $D_{1,D}$ and the threshold value used by the referee $D_{1,th}$. As before, $Q$ is the upper bound on the quantum transmitted information, $\gamma$ is the quantum advantage and $\epsilon$ the error probability of the protocol.}\label{Tab:results}
\end{table*}

The above theorem guarantees that, if we sacrifice an arbitrarily small quantity $\epsilon$ of the rate with respect to the Gilbert-Varshamov bound (i.e., we set $R= 1-H_2(\delta)-\epsilon$), the probability of obtaining an incorrect minimum distance decreases exponentially with the output size. For example, for a value of $m=10^7$ and $\epsilon=10^{-3}$, this probability is less than $10^{-10^4}$.

\begin{thm}\label{thm:rank}
Let $G \in \mathbb{F}_2^{n \times m}$ be a Toeplitz matrix
chosen uniformly at random. Then,
\[
\Pr( G \textrm{ is not full rank})=2^{-1} 2^{-m(1-R)}.
\]
\end{thm}

Theorem~\ref{thm:rank} is an immediate consequence of
Theorem~1 in \cite{rank}. Once again, this probability decreases exponentially with the output size $m$.\\

\subsection*{Detailed Experimental Results}

In Table~\ref{Tab:results}, we report the complete results of our experiment. The dominating source of uncertainty is the uncertainty in the total mean photon number of the signals. This uncertainty is due to the summation of the fluctuations of several devices, such as laser power, VOA, and varying loss in the channel. For each input size $n$, we perform a calibration process to determine $\mu$. In this process, with a proper value of VOA selected from our numerical optimization, the referee sends out around $10^7\sim10^8$ pulses to Alice and Bob. From the total detection counts on $D_0$ and $D_1$ and the pre-calibrated losses (Table~\ref{Tab:parameters}), we estimate the $\mu$. We repeat this calibration process a few rounds and obtain the mean value and the standard deviation for $\mu$. These results are shown in the second column of Table \ref{Tab:results}. For all tested cases, the uncertainty in mean photon number was below $4\%$.

From our model of the protocol, we use the uncertainty in the mean photon number to directly calculate an uncertainty for the quantum transmitted information as well as for the error probability of the protocol. As it can be seen from Table \ref{Tab:results}, all error probabilities are compatible with the system operating below the target value of $\epsilon=5\times10^{-5}$. Additionally, we have included the average values observed for the number of clicks in detector $D_1$ for equal and different inputs, as well as the threshold values used by the referee.

Finally, we estimate the effect of detector dead times in our experiment as follows. For each input size, we can calculate the probability $p$ that an individual pulse leads to a click in detector $D_1$. In our setup, after a click occurs, the following 50 pulses are blocked by the detector and cannot be registered. The probability $p'$ that a click occurs for these 50 pulses is given by $p'=1-(1-p)^{50}\approx 50p$. This number is very small whenever $p$ is small, as is the case in our experiment. For instance, for an input size of $n=1.42\times 10^8$, the expected number of blocked clicks is approximately $0.1\%$ of the total expected clicks. Therefore, this effect is negligible compared to fluctuations in the mean photon number, which is of the order of $4\%$.

\bibliography{References}

\begin{thebibliography}{10}

\bibitem{bennett84a}
C.~H. Bennett and G.~Brassard, ``Quantum cryptography: Public key distribution
  and coin tossing.,'' in {\em Proceedings of IEEE International Conference on
  Computers, Systems, and Signal Processing, Bangalore, India}, (New York),
  pp.~175--179, IEEE, dec 1984.

\bibitem{ekert91a}
A.~Ekert, ``Quantum cryptography based on bell's theorem,'' {\em Phys. Rev.
  Lett.}, vol.~67, no.~6, pp.~661--663, 1991.

\bibitem{QDS}
D.~Gottesman and I.~Chuang, ``Quantum digital signatures,'' {\em arXiv preprint
  quant-ph/0105032}, 2001.

\bibitem{giovannetti2011advances}
V.~Giovannetti, S.~Lloyd, and L.~Maccone, ``Advances in quantum metrology,''
  {\em Nature Photonics}, vol.~5, no.~4, pp.~222--229, 2011.

\bibitem{Yao1979}
A.~C.-C. Yao, ``Some complexity questions related to distributive computing
  (preliminary report),'' in {\em Proceedings of the 11th Annual ACM Symposium
  on Theory of Computing}, pp.~209--213, 1979.

\bibitem{BrassardQCC}
G.~Brassard, ``Quantum communication complexity,'' {\em Foundations of
  Physics}, vol.~33, no.~11, pp.~1593--1616, 2003.

\bibitem{RevModPhys.82.665}
H.~Buhrman, R.~Cleve, S.~Massar, and R.~de~Wolf, ``Nonlocality and
  communication complexity,'' {\em Rev. Mod. Phys.}, vol.~82, pp.~665--698, Mar
  2010.

\bibitem{buhrman1999multiparty}
H.~Buhrman, W.~van Dam, P.~H{\o}yer, and A.~Tapp, ``Multiparty quantum
  communication complexity,'' {\em Physical Review A}, vol.~60, no.~4, p.~2737,
  1999.

\bibitem{buhrman1998quantum}
H.~Buhrman, R.~Cleve, and A.~Wigderson, ``Quantum vs. classical communication
  and computation,'' in {\em Proceedings of the thirtieth annual ACM symposium
  on Theory of computing}, pp.~63--68, ACM, 1998.

\bibitem{QuantumFingerprinting}
H.~Buhrman, R.~Cleve, J.~Watrous, and R.~de~Wolf, ``Quantum fingerprinting,''
  {\em Phys. Rev. Lett.}, vol.~87, p.~167902, Sep 2001.

\bibitem{RazProblem}
R.~Raz, ``Exponential separation of quantum and classical communication
  complexity,'' in {\em Proceedings of the Thirty-First Annual ACM Symposium on
  Theory of Computing}, pp.~358--367, 1999.

\bibitem{HM-Bar-Yossef}
Z.~Bar-Yossef, T.~S. Jayram, and I.~Kerenidis, ``Exponential separation of
  quantum and classical one-way communication complexity,'' in {\em Proceedings
  of the Thirty-Sixth Annual ACM Symposium on Theory of Computing},
  pp.~128--137, 2004.

\bibitem{gavinsky2007exponential}
D.~Gavinsky, J.~Kempe, I.~Kerenidis, R.~Raz, and R.~De~Wolf, ``Exponential
  separations for one-way quantum communication complexity, with applications
  to cryptography,'' in {\em Proceedings of the thirty-ninth annual ACM
  symposium on Theory of computing}, pp.~516--525, 2007.

\bibitem{regev2011quantum}
O.~Regev and B.~Klartag, ``Quantum one-way communication can be exponentially
  stronger than classical communication,'' in {\em Proceedings of the
  forty-third annual ACM symposium on Theory of computing}, pp.~31--40, 2011.

\bibitem{becerra2013experimental}
F.~Becerra, J.~Fan, G.~Baumgartner, J.~Goldhar, J.~Kosloski, and A.~Migdall,
  ``Experimental demonstration of a receiver beating the standard quantum limit
  for multiple nonorthogonal state discrimination,'' {\em Nature Photonics},
  vol.~7, no.~2, pp.~147--152, 2013.

\bibitem{xiang2011entanglement}
G.-Y. Xiang, B.~L. Higgins, D.~Berry, H.~M. Wiseman, and G.~Pryde,
  ``Entanglement-enhanced measurement of a completely unknown optical phase,''
  {\em Nature Photonics}, vol.~5, no.~1, pp.~43--47, 2011.

\bibitem{ng2012exp}
N.~H.~Y. Ng, S.~K. Joshi, C.~C. Ming, C.~Kurtsiefer, and S.~Wehner,
  ``Experimental implementation of bit commitment in the noisy-storage model,''
  {\em Nature communications}, vol.~3, p.~1326, 2012.

\bibitem{clarke2012experimental}
P.~J. Clarke, R.~J. Collins, V.~Dunjko, E.~Andersson, J.~Jeffers, and G.~S.
  Buller, ``Experimental demonstration of quantum digital signatures using
  phase-encoded coherent states of light,'' {\em Nature communications},
  vol.~3, p.~1174, 2012.

\bibitem{lunghi2013exp}
T.~Lunghi, J.~Kaniewski, F.~Bussi{\`e}res, R.~Houlmann, M.~Tomamichel, A.~Kent,
  N.~Gisin, S.~Wehner, and H.~Zbinden, ``Experimental bit commitment based on
  quantum communication and special relativity,'' {\em Phys. Rev. Lett.},
  vol.~111, no.~18, p.~180504, 2013.

\bibitem{liu2014exp}
Y.~Liu, Y.~Cao, M.~Curty, S.-K. Liao, J.~Wang, K.~Cui, Y.-H. Li, Z.-H. Lin,
  Q.-C. Sun, D.-D. Li, {\em et~al.}, ``Experimental unconditionally secure bit
  commitment,'' {\em Phys. Rev. Lett.}, vol.~112, no.~1, p.~010504, 2014.

\bibitem{collins2014realization}
R.~J. Collins, R.~J. Donaldson, V.~Dunjko, P.~Wallden, P.~J. Clarke,
  E.~Andersson, J.~Jeffers, and G.~S. Buller, ``Realization of quantum digital
  signatures without the requirement of quantum memory,'' {\em Phys. Rev.
  Lett.}, vol.~113, no.~4, p.~040502, 2014.

\bibitem{berlin2011exp}
G.~Berl{\'\i}n, G.~Brassard, F.~Bussi{\'e}res, N.~Godbout, J.~A. Slater, and
  W.~Tittel, ``Experimental loss-tolerant quantum coin flipping,'' {\em Nature
  communications}, vol.~2, p.~561, 2011.

\bibitem{pappa2014experimental}
A.~Pappa, P.~Jouguet, T.~Lawson, A.~Chailloux, M.~Legr{\'e}, P.~Trinkler,
  I.~Kerenidis, and E.~Diamanti, ``Experimental plug and play quantum coin
  flipping,'' {\em Nature communications}, vol.~5, p.~3717, 2014.

\bibitem{kushilevitz2006communication}
E.~Kushilevitz and N.~Nisan, {\em Communication Complexity}.
\newblock Cambridge University Press, 2006.

\bibitem{HornFPs}
R.~T. Horn, S.~A. Babichev, K.-P. Marzlin, A.~I. Lvovsky, and B.~C. Sanders,
  ``Single-qubit optical quantum fingerprinting,'' {\em Phys. Rev. Lett.},
  vol.~95, p.~150502, Oct 2005.

\bibitem{du2006experimental}
J.~Du, P.~Zou, X.~Peng, D.~K. Oi, L.~Kwek, C.~Oh, and A.~Ekert, ``Experimental
  quantum multimeter and one-qubit fingerprinting,'' {\em Phys. Rev. A},
  vol.~74, no.~4, p.~042319, 2006.

\bibitem{TrojekQCC}
P.~Trojek, C.~Schmid, M.~Bourennane, C.~Brukner, M.~Zukowski, and
  H.~Weinfurter, ``Experimental quantum communication complexity,'' {\em Phys.
  Rev. A}, vol.~72, p.~050305, Nov 2005.

\bibitem{arrazolaqfp}
J.~M. Arrazola and N.~L\"utkenhaus, ``Quantum fingerprinting with coherent
  states and a constant mean number of photons,'' {\em Phys. Rev. A}, vol.~89,
  p.~062305, Jun 2014.

\bibitem{arrazola2014QC}
J.~M. Arrazola and N.~L{\"u}tkenhaus, ``Quantum communication with coherent
  states and linear optics,'' {\em Phys. Rev. A}, vol.~90, no.~4, p.~042335,
  2014.

\bibitem{arrazola2014TQC}
J.~M. Arrazola and N.~L{\"u}tkenhaus, ``Quantum communication complexity with
  coherent states and linear optics,'' in {\em 9th Conference on the Theory of
  Quantum Computation, Communication and Cryptography}, p.~36, 2014.

\bibitem{massar2005quantum}
S.~Massar, ``Quantum fingerprinting with a single particle,'' {\em Physical
  Review A}, vol.~71, no.~1, p.~012310, 2005.

\bibitem{babai1997randomized}
L.~Babai and P.~G. Kimmel, ``Randomized simultaneous messages: Solution of a
  problem of yao in communication complexity,'' in {\em Proceedings of the 12th
  Annual IEEE Conference on Computational Complexity}, pp.~239--246, IEEE, IEE,
  Los Alamitos, California, 1997.

\bibitem{newman1996public}
I.~Newman and M.~Szegedy, ``Public vs. private coin flips in one round
  communication games,'' in {\em Proceedings of the twenty-eighth annual ACM
  symposium on Theory of computing}, pp.~561--570, 1996.

\bibitem{idquantique}
IDQuantique, Geneva, http://www.idquantique.com.

\bibitem{Gilbert}
E.~N. Gilbert, ``A comparison of signalling alphabets,'' {\em Bell System
  Technical Journal}, vol.~31, no.~3, pp.~504--522, 1952.

\bibitem{Varshamov}
R.~Varshamov, ``Estimate of the number of signals in error correcting codes,''
  in {\em Dokl. Akad. Nauk SSSR}, vol.~117, pp.~739--741, 1957.

\bibitem{barg2002random}
A.~Barg and G.~Forney, ``Random codes: Minimum distances and error exponents,''
  {\em IEEE Transactions on Information Theory}, vol.~48, no.~9,
  pp.~2568--2573, 2002.

\bibitem{handbook}
I.~Z. Emiris and V.~Y. Pan, ``Applications of fft and structured matrices,'' in
  {\em Algorithms and theory of computation handbook}, pp.~18--18, Chapman \&
  Hall/CRC, 2010.

\bibitem{frigo2005design}
M.~Frigo and S.~G. Johnson, ``The design and implementation of fftw3,'' {\em
  Proceedings of the IEEE}, vol.~93, no.~2, pp.~216--231, 2005.

\bibitem{xu2012ultrafast}
F.~Xu, B.~Qi, X.~Ma, H.~Xu, H.~Zheng, and H.-K. Lo, ``Ultrafast quantum random
  number generation based on quantum phase fluctuations,'' {\em Optics
  express}, vol.~20, no.~11, pp.~12366--12377, 2012.

\bibitem{stucki02a}
D.~Stucki, N.~Gisin, O.~Guinnard, G.~Ribordy, and H.~Zbinden, ``Quantum key
  distribution over 67 km with a plug\&play system,'' {\em New J. Phys.},
  vol.~4, p.~41, 2002.

\bibitem{gavinsky2013secrets}
D.~Gavinsky and T.~Ito, ``Quantum fingerprints that keep secrets,'' {\em
  Quantum Information \& Computation}, vol.~13, no.~7-8, pp.~583--606, 2013.

\bibitem{chakrabarti2001informational}
A.~Chakrabarti, Y.~Shi, A.~Wirth, and A.~Yao, ``Informational complexity and
  the direct sum problem for simultaneous message complexity,'' in {\em
  Foundations of Computer Science, 2001. Proceedings. 42nd IEEE Symposium on},
  pp.~270--278, IEEE, 2001.

\bibitem{marsili2013detecting}
F.~Marsili, V.~Verma, J.~Stern, S.~Harrington, A.~Lita, T.~Gerrits,
  I.~Vayshenker, B.~Baek, M.~Shaw, R.~Mirin, {\em et~al.}, ``Detecting single
  infrared photons with 93\% system efficiency,'' {\em Nature Photonics},
  vol.~7, no.~3, pp.~210--214, 2013.

\bibitem{essentialcoding}
V.~Guruswami, A.~Rudra, and M.~Sudan, {\em Essential Coding Theory}.
\newblock University of Buffalo, 2014.

\bibitem{rank}
D.~Daykin, ``Distribution of bordered persymmetric matrices in a finite
  field,'' {\em J. Reine Angew. Math.(Crelle’s J.)}, vol.~203, pp.~47--54,
  1960.

\end{thebibliography}
\bibliographystyle{ieeetr}
\end{document}